\documentclass[twocolumn,showpacs,preprintnumbers,amsmath,amssymb, superscriptaddress, prb]{revtex4-1}

\usepackage{stmaryrd}
\usepackage{txfonts}
\usepackage{amssymb}
\usepackage{mathrsfs}
\usepackage{graphicx}
\usepackage{dcolumn}
\usepackage{bm}
\usepackage{epsfig}
\usepackage{color}                    
\usepackage{hyperref}                 

\begin{document}
\preprint{\href{http://dx.doi.org/10.1103/PhysRevB.86.054506}{S. Z. Lin and X. Hu, Phys. Ref. {\bf{B}} {\bf 86}, 054506 (2012).}}
\title{In-plane dissipation as a possible synchronization mechanism for terahertz radiation from intrinsic Josephson junctions of layered superconductors}
\author{Shi-Zeng Lin}
\email{szl@lanl.gov}
\affiliation{Theoretical Division, Los Alamos National Laboratory, Los Alamos, New Mexico 87545, USA}

\author{Xiao Hu}
\affiliation{International Center for Materials Nanoarchitectonics (WPI-MANA), National Institute for Materials Science, Tsukuba 305-0044, Japan}

\begin{abstract}
Strong terahertz radiation from mesa structure of $\rm{Bi_2Sr_2CaCu_2O_{8+\delta}}$ single crystal has been observed recently, where the mesa intrinsically forms a cavity. For a thick mesa of large number of junctions, there are many cavity modes with different wave vectors along the $c$-axis corresponding to almost the same bias voltages. The mechanism responsible for exciting the uniform mode which radiates coherent terahertz waves in experiments is unknown. In this work, we show that the in-plane dissipation selects the uniform mode. For perturbations with non-zero wave numbers along the $c$-axis, the in-plane dissipations are significantly enhanced, which prevent the excitation of corresponding cavity modes. Our analytical results are confirmed by numerical simulations.
\end{abstract}

\pacs{74.50.+r, 74.25.Gz, 85.25.Cp}

\date{\today}

 \maketitle
\section { Introduction }
 A layered cuprate superconductor, such as $\rm{Bi_2Sr_2CaCu_2O_{8+\delta}}$ (BSCCO), intrinsically forms a stack of Josephson junctions\cite{Kleiner92}. Because of the large superconducting energy gap (60 meV), these build-in intrinsic Josephson junctions (IJJs) can support oscillations with frequencies in the terahertz (THz) band. IJJs are homogeneous and packed on nanometer
scale, much smaller than THz electromagnetic (EM) wavelength. If synchronized, they can radiate powerful THz EM waves. The radiation frequency is determined by the bias voltage according to the ac Josephson relation, thus in principle can be tuned continuously. The THz generator based on IJJs thus is promising to fill the THz gap\cite{Hu10,Savelev10}.

In 2007, coherent radiations from a mesa structure of BSCCO without external magnetic fields were detected experimentally\cite{Ozyuzer07}. The radiation frequency and voltage follow the ac Josephson relation, thus the radiation is due to the Josephson plasma oscillation in the mesa. The frequency $f$ is determined by lateral size $L_x$ of mesa $f=c_0/(2L_x)$ with $c_0=c/\sqrt{\epsilon_c}$ the Josephson plasma velocity where $\epsilon_c$ is the dielectric constant of BSCCO. The measured relation between $L_x$ and $f$ has revealed unambiguously that the mesa works as a cavity to synchronize plasma oscillations in different junctions. The cavity resonance mechanism has been confirmed by many independent experiments\cite{kadowaki08,Wang09,Wang10,Tsujimoto10,Tsujimoto12b}.

The experiments raise several questions. First in experiments a dc current is uniformly injected into the mesa. One thus expects the superconducting phase would oscillate homogeneously along the lateral directions, which, however, seemed to be hardly reconciled with the observed cavity modes. This question has been addressed in Refs. \onlinecite{szlin08b,Koshelev08b}. They suggested that superconducting phase develops $\pi$ phase kinks near the cavity resonances. With the help of the $\pi$ phase kink, the standing wave of electromagnetic fields can be stabilized and a large amount of energy is pumped into EM waves from the dc current.

Secondly, as the mesa intrinsically forms a three-dimensional (3D) cavity, cavity modes both along the $c$-axis and $ab$-plane can be excited. When the frequency of the plasma oscillation $\omega$ determined by the bias voltage per junction is tuned to the cavity frequency  $\omega_c'=\sqrt{(m_x\pi/L_x)^2+(m_y\pi/L_y)^2} c_{q}$, the cavity mode $(m_x, m_y)$ is excited\cite{Kleiner94}, where $c_q$ depends on the wave vector $q=n \pi/(N+1)$ along the $c$-axis with $N$ the number of junctions and $n$ an integer. For a large $N$, such as $N\approx 1000$ in experiments, there are many cavity modes with different $q$ within a narrow window of bias voltage, and one would expect various modes should be excited when the voltage is swept. Nevertheless, for a given radiating sample, only the modes uniform along the $c$-axis ($q=0$) is observed upon sweeping current in experiment. The reason remains illusive.

Thirdly, understanding on possible cavity modes along the $c$-axis is also important for getting stronger radiation from the mesa structure of BSCCO, which is still too weak for practical applications to date. The radiation power can be enhanced by using thicker mesa with larger $N$, since the radiation power is proportional to $N^2$ in the superradiation region. One question is whether there exists a fundamental limiting factor besides the heating effect for $N$.

In this paper, we show analytically that in-plane dissipations prevent the excitation of non-uniform cavity modes along the $c$-axis. For non-uniform perturbations with a large $q$, effective in-plane dissipations are greatly enhanced, and thus the perturbations quickly die out and no cavity mode is excited. For perturbations with a small $q$, in-plane dissipations are weak and vanish for $q=0$. The weak dissipation along the $c$-axis cannot damp them efficiently, and thus cavity modes with small $q$'s are excited. For $N\approx 10^3$, only the cavity mode with $q=0$ can be excited. While for large $N\approx 10^4$, modes with finite but small $q$ vectors can also be excited and compete with mode $q=0$. The analytical results are confirmed by direct numerical calculations. The mechanism to achieve synchronization by dissipation may be applied to other systems as well by preventing the excitation of the out-of-phase mode.

\section{ Model } 
We consider a stack of IJJs with lateral sizes $L_x\approx 80 \rm{\ \mu m}$, $L_y\approx 300 \rm{\ \mu m}$ similar to those in experiments, see the inset of Fig. \ref{f1}. Because $L_y\gg L_x$, we can assume that the superconducting phase is uniform along the $y$-axis. The dynamics of the gauge invariant phase difference $\varphi_l$ and magnetic field $B_{y,l}$ in the $l$-th junction are described by \cite{Sakai93,Bulaevskii94,Bulaevskii96,Machida99,Koshelev01}
\begin{equation}\label{eq1}
\partial _t^2\varphi _l+\beta _c \partial _t\varphi _l+\text{sin$\varphi $}_l=\partial _xB_{y,l},
\end{equation}
\begin{equation}\label{eq2}
\left[\zeta  \Delta ^{(2)}-\left(1+\beta _{\text{ab}}\partial _t\right)\right]B_{y,l}+\left(1+\beta _{\text{ab}}\partial _t\right)\partial _x\varphi _l=0,
\end{equation}
where $\Delta^{(2)}f_l \equiv f_{l+1}+f_{l-1}-2 f_l$ is the finite difference operator. $\zeta\approx 10^5$ is the inductive coupling, $\beta_c\approx 0.02$ and $\beta_{ab}\approx 0.2$ are the renormalized conductivity along the $c$ axis and $ab$ plane respectively.\cite{units} For BSCCO, it is well established that the in-plane conductivity $\sigma_{ab}$ is much larger than the $c$-axis conductivity $\sigma_{c}$, from microwave \cite{Lee96}, infrared spectroscopy\cite{Romero92} and transport \cite{Latyshev03} measurements. At temperature $T=0$, $\sigma_{ab}\approx 4\times 10^6\rm{\ (\Omega\cdot m)^{-1}}$ and $\sigma_{c}\approx 0.2 \rm{\ (\Omega\cdot m)^{-1}}$. Both $\beta_c$ and $\beta_{ab}$ depend on frequency, but the dependence is weak in the interested frequency region\cite{Latyshev1999,Corson2000}. In the following discussion we will neglect the frequency dependence. The generalization is straightforward as we are
working in the frequency domain. The in-plane dissipation due to $\beta_{ab}$ has been overlooked in many theoretical models.

Equations (\ref{eq1}) and (\ref{eq2}) are supplemented by the boundary conditions. When the phase oscillates uniformly along the $c$-axis, strong radiation of EM waves occurs, which can be accounted for using the boundary condition\cite{Bulaevskii06PRL,Bulaevskii07}
\begin{equation}\label{eq3}
{B_y}(\omega )=\mp\frac{E_z(\omega )}{Z(\omega )},\ \ \ \  Z=\frac{2}{\sqrt{\varepsilon _d}L_z\left[\left|k_{\omega }\right|\text{}-\frac{2i}{\pi }k_{\omega }\ln \frac{5.03}{\left|k_{\omega }\right|L_z}\right]},
\end{equation}
where $-$ ($+$) corresponds to the edge $x=L_x$ ($x=0$), and $k_{\omega}=\omega\sqrt{\epsilon_d}$ with $\epsilon_d$ the dielectric constant of the dielectric medium outside the IJJs. For stacks with height $L_z\ll 100\rm{\ \mu m}$, $Z\gg 1$. For non-uniform oscillations along the $c$-axis, the radiation is weak and we can use the non-radiating boundary condition $B_{y,l}=\pm I_{\rm{ext}} L_x/2$ with $I_{\rm{ext}}$ the bias current. We assume that the IJJs stack is sandwiched by two good conductors, such that the tangential current inside the conductor is zero. We use the boundary condition $B_{y, l=1}=B_{y,l=0}$ and similarly for $l=N$, which corresponds to $\partial_z B_y(z)=0$ in the continuum limit.

\section {Instability of the homogeneous solution} 
In experiments, one first ramps up the bias current and when current exceeds the critical one, the IJJs switch into the resistive state. One then reduces the current to the target value where radiation is observed. In the resistive state with the current close to the critical one, the phase $\varphi_l$ oscillates homogeneous along the $x$ direction. The phases may be either uniform along the $c$-axis or different in different junctions. Here we show that the solution with $\varphi_l$ homogeneous along the $x$ direction is unstable when the bias current is reduced and that only the cavity modes with a long wavelength $q\ll 1$ along the $c$-axis can be excited.

For the uniform solution along the $c$-axis $\varphi_l=\varphi_0$, the solution to Eqs. (\ref{eq1}) and (\ref{eq2}) in the THz frequency region $\omega\gg 1$ can be written as $\varphi _0=\omega  t +g(x) \exp (i \omega  t)$ with 
\begin{equation}\label{eq5}
g(x)=\frac{1}{-\omega ^2+i \beta_c  \omega }\left[i-\frac{\cos\left[\left(x-L_x/2\right) \omega \right]}{Z \sin\left({L_x \omega }/{2}\right)-i \cos\left({L_x \omega }/{2}\right)}\right].
\end{equation}
The first term in the square bracket of Eq. (\ref{eq5}) is due to plasma oscillation and the second term is due to radiation. The frequency $\omega$ can be tuned by the bias current $I_{\rm{ext}}$ (or voltage).

 \begin{figure}[t]
\psfig{figure=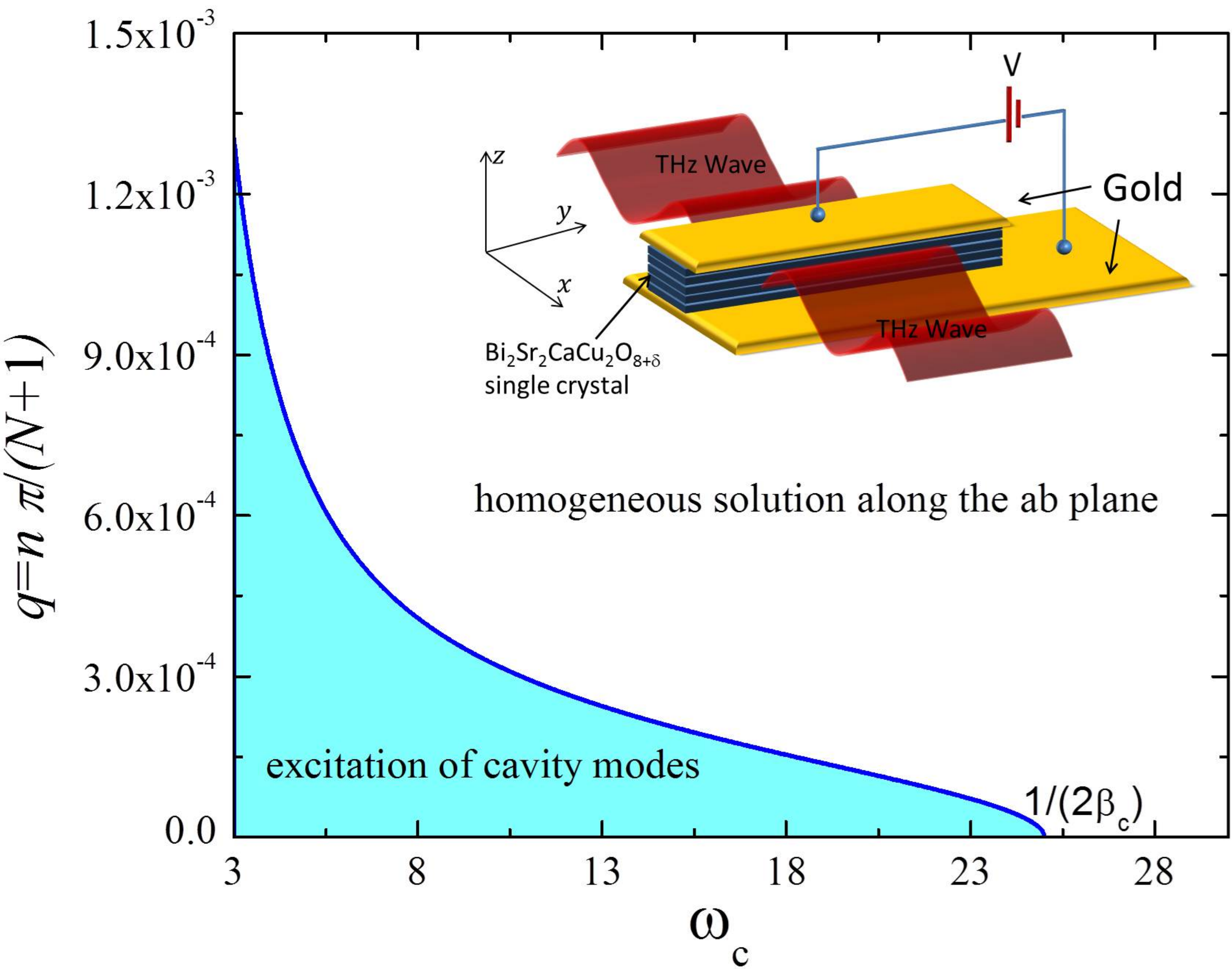,width=\columnwidth}
\caption{\label{f1}(color online) Stability diagram for the homogeneous solution along the $ab$ plane. In the filled region, the homogeneous solution is unstable and the cavity mode $(k_m, q)$ is excited. Inset is a schematic view of the setup for THz radiation.}
\end{figure}

To reveal the instability of the homogeneous solution, we add small perturbations to the uniform solution $\varphi _l=\varphi_0+\theta _l$, $B_{y,l}=B_0+\tilde{B}_{y,l}$ with $B_0=\partial_x\varphi_0$, $\theta_l\ll 1$ and $\tilde{B}_{y,l}\ll 1$. Since the perturbations are non-uniform along the $c$-axis, the radiation contribution can be neglected, thus we can use the non-radiating boundary condition, $\partial _x\theta _l=0$ and $\tilde{B}_{y,l}=0$. The solution for $\theta_l$ and $\tilde{B}_{y,l}$ can be written as
\begin{equation}\label{eq6}
\theta _l(x,t)=\sum _{m,q,p}\cos ( q l)\cos \left(k_mx\right)a_p(m,q) \exp [i (p \omega-\Omega)  t]
\end{equation}
\begin{equation}\label{eq7}
\tilde{B}_{y,l}(x,t)=\sum _{m,q,p}\cos ( q l)\sin \left(k_mx\right) b_p(m,q) \exp [i (p \omega-\Omega)  t]
\end{equation}
with $k_m=m \pi /L_x$, $q=n\pi/(N+1)$ and $p$ an integer. The perturbations with frequency $\Omega$ couple with the nonlinear Josephson current $\sin\varphi_l$ and induce frequency harmonics $p\omega-\Omega$.\cite{Lin11b} From Eq. (\ref{eq2}), we have $b_p=-c_q^2 k_m  a_p$ with the plasma velocity $c_{q}(\omega)=\left[1+2\zeta  (1-\cos  q)/(1+\beta _{ab} i \omega)\right]^{-1/2}$. Substituting Eqs. (\ref{eq6}) and (\ref{eq7}) into Eqs. (\ref{eq1}) and (\ref{eq2}), and comparing each frequency component, we obtain equation for perturbations
\[
\left[ \omega _m^2(\omega_p) -\omega_p^2+i \omega_p \beta _c\right]a_p+\frac{a_{p-1}+a_{p+1}}{2}-\frac{A_{p-2}-A_p}{2i}=0
\]
where $\omega _m^2=c_{q}^2k_m^2$, $\omega_p=p \omega -\Omega$, and $A_p=\frac{2}{L_x}\int _0^{L_x}dx g(x) \theta _p(x) \cos(k_m x)$. Because of the in-plane damping $\beta_{ab}$, $\omega_m^2$ is a complex number. We can split $\omega_m^2$ into real and imaginary parts $\omega _m^2(\omega )=\Omega _m^2(\omega )+i \mathcal{B}_{\text{ab}}(\omega ) \omega$, with
\begin{equation}\label{eq10}
\Omega _m^2(\omega )=\frac{1+2\zeta  (1-\cos  q)+\beta _{\text{ab}}^2\omega ^2}{[1+2\zeta  (1-\cos  q)]^2+\beta _{\text{ab}}^2\omega ^2}k_m^2,
\end{equation}
\begin{equation}\label{eq11}
\mathcal{B}_{\text{ab}}(\omega ) =\frac{2\zeta  (1-\cos  q)\beta _{\text{ab}}}{[1+2\zeta  (1-\cos  q)]^2+\beta _{\text{ab}}^2\omega ^2}k_m^2.
\end{equation}
$\Omega _m(\omega)$ is the cavity resonance frequency for the mode $(k_m, q)$. For the non-uniform perturbations along the $c$-axis $q>0$, the effective in-plane damping is enhanced according to $\mathcal{B}_{\text{ab}}$. It is this enhanced in-plane dissipation that prevents the excitation of non-uniform cavity mode along the $c$-axis with a large $q$ as revealed later.

In the region of $\omega_m\gg 1$ and $\beta_c\ll 1$, we have $|\Omega_m^2-\omega_{p=0}^2|\ll 1$. Thus the dominant wave vector of $\theta _p(x)$ is $k_m$. The dominant wave vector for $g(x)$ is $k_x=0$ because the radiation contribution is small. Then we can approximate $\theta _p(x)g(x)$ by $\bar{g}\theta _p\left(k_m\right)\cos(k_m x)$ with $\bar{g}$ the spatial average of $g(x)$, because other modes are negligibly small. Neglecting the small dissipation contribution $\beta _c$ in $\bar{g}$, we have
\[
\frac{\bar{g}}{2i}=\frac{1}{-2\omega ^2}+\frac{1}{L_x \omega ^3 \left[\cot \left({L_x \omega }/{2}\right)+Z i \right]}=\frac{1}{-2 \omega ^2}+R_r+i R_i
\]
where $R_r$ is the real part of the radiation contribution and $R_i$ is the imaginary part. We then have $A_p=\bar{g}a_p$. The equation for perturbations then can be written as
\begin{align}\label{eq13}
\nonumber \left[ \Omega _m^2 -\omega_p^2+R_r-\frac{1}{2\omega^2}\right]a_p+i \left[\omega_p \left(\beta _c+\mathcal{B}_{\text{ab}}\right)+R_i\right] a_p\\
+\frac{1}{2}\left(a_{p-1}+a_{p+1}\right)+\frac{1}{2 \omega ^2}a_{p-2}-R_ra_{p-2}-i R_ia_{p-2}=0.
\end{align}
The radiation shifts the resonance frequency by $R_r$ and also contribution to the damping through $R_i$. Both $R_r$ and $R_i$ have the order $1/\omega^2\ll 1$ for $L_x\sim 1$. The resonant frequency for perturbations with a wave vector $(k_m, q)$ is $\Omega=\omega_c$ with $\omega_c$ given by $\omega_c^2=\Omega_m^2(\omega_c)+R_r(\omega)-\frac{1}{2\omega^2}$. $\omega_c$ is also the cavity frequency for the cavity mode $(k_m, q)$.

We then show that the parametric instability develops at a voltage when $\omega= 2\omega_c+\delta$ with $\delta\ll 1$. As $\Omega\approx \Omega_m\approx \omega_c$, the dominant frequency components are $p=0$ and $1$. The other frequency harmonics ($p>1$ or $p<0$) are small because their amplitude is of order $\left([(2p-1)^2-1]\Omega_m^2\right)^{-1} a_{0, 1}\ll a_{0,1}$ thus can be neglected. Then Eq. (\ref{eq13}) can be written as
\begin{equation}\label{eq15}
\left[-2\omega _c\Omega _{\delta }+i \left(-\omega _c\left(\beta _c+\mathcal{B}_{\text{ab}}\right)+R_i\right)\right]a_0+a_1/2=0,
\end{equation}
\begin{equation}\label{eq16}
\left[-2\omega _c\left(\delta -\Omega _{\delta }\right)+i \left(\omega _c\left(\beta _c+\mathcal{B}_{\text{ab}}\right)+R_i\right)\right]a_1+a_0/2=0,
\end{equation}
with $\Omega =\Omega _{\delta }+\omega _c$ and $\Omega _{\delta }\ll 1$. The spectrum of the perturbations $\Omega$ is given by equating the determinant of the coefficients matrix in Eqs. (\ref{eq15}) and (\ref{eq16}) to zero, which yields
\[
\Omega _{\delta }=\left(-\frac{i \mathcal{B}_{\text{ab}}}{2}-\frac{i \beta _c}{2}+\frac{\delta }{2}\right)\pm \left(\frac{i}{4 \omega _c}-\frac{i}{2} \omega _c \delta ^2-\delta  R_i+\frac{i}{2 \omega _c}R_i^2\right)
\]
The homogeneous solution along the $x$ direction is unstable when ${\rm{Im}}[\Omega _{\delta }]>0$, which gives
\begin{equation}\label{eq18}
\delta ^2<-\frac{ \left(\mathcal{B}_{\text{ab}}\left(\omega _c\right)+\beta _c\right)}{\omega _c}+\frac{1}{2 \omega _c^2}+\frac{1}{ \omega _c^2}R_i^2.
\end{equation}
From this expression, it becomes clear that the radiation tends to destabilize the homogeneous solution, albeit with a small impact $R_i\sim 1/\omega^2 \ll 1$. More importantly, both the effective in-plane dissipation $\mathcal{B}_{ab}$ and dissipation along the $c$-axis $\beta_c$ tend to stabilize the solution homogeneous along the $x$ axis. For long wavelength perturbations along the $c$-axis, $\zeta q^2\ll 1$, the dissipation is weak $\mathcal{B}_{\text{ab}}\left(\omega _c\right)+\beta _c<1/(2\omega_c)$ thus the cavity mode $(k_m, q)$ can be excited. By using Eqs. (\ref{eq10}), (\ref{eq11}) and (\ref{eq18}), the condition for the excitation of the mode $(k_m, q)$ for $\beta_{ab}>0$ is
\begin{equation}\label{eq19}
q^2<-\frac{1}{\zeta}{\left(\beta _c-\frac{1}{2\omega _c}\right)\left(1+\beta _{\text{ab}}^2\omega _c^2\right)}{ \left(\omega _c^2\beta _{\text{ab}}+\beta _c-\frac{1}{2\omega _c}\right)^{-1}}.
\end{equation}
The stability diagram is presented in Fig. \ref{f1} with the given parameters. Only the cavity mode with long wavelength along the $c$-axis can be excited. As $q=n\pi/(N+1)$, for $N\approx 10^3$ used in experiments, only the uniform cavity $n=0$ can be excited. For a larger $N>10^4$, modes with finite but small $n$ can also be excited.

\begin{figure*}[t]
\psfig{figure=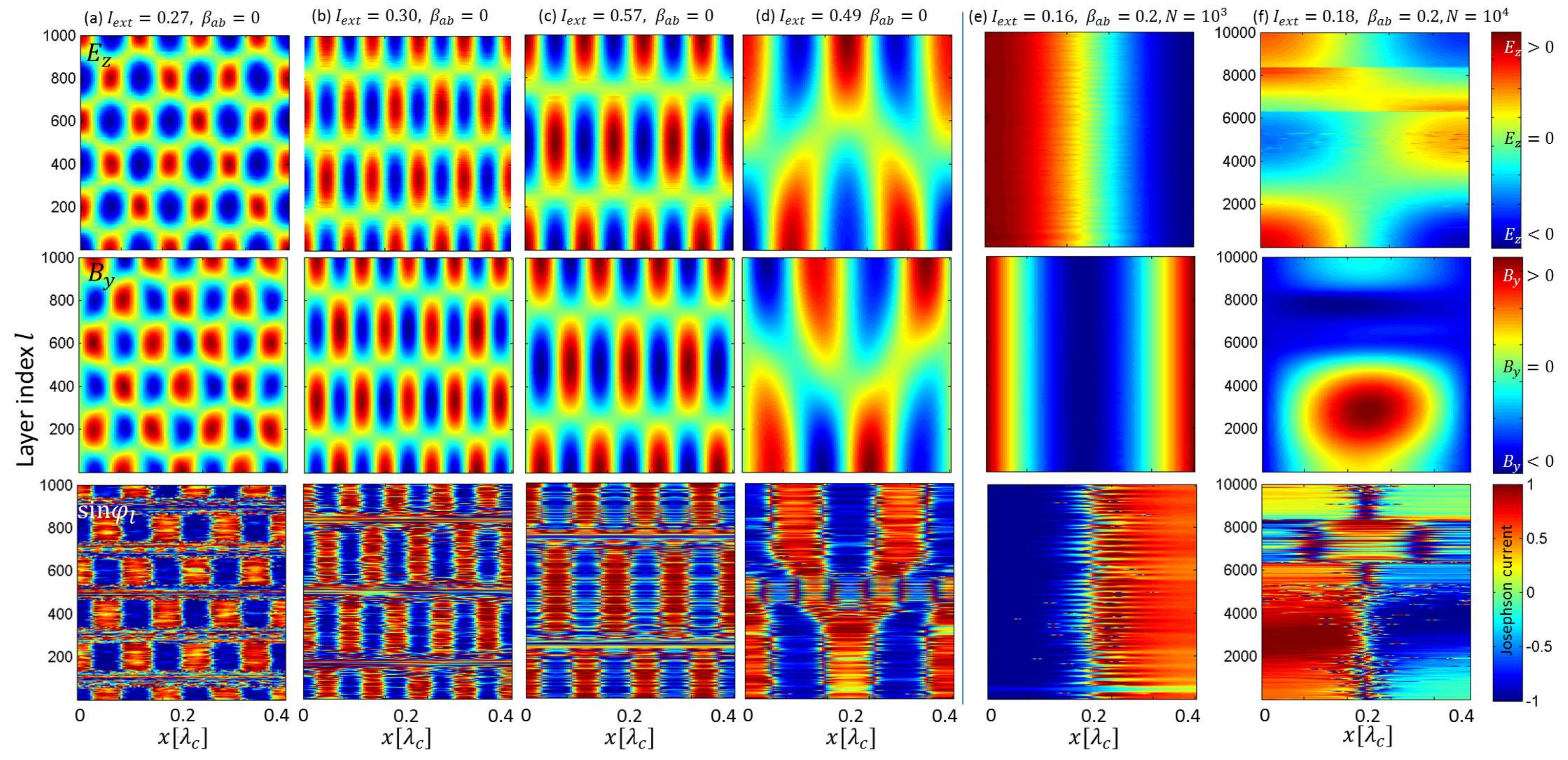,width=18cm}\label{f2}
\caption{\label{f2}(color online) Snapshots of the electric field (first row), magnetic field (second row) and Josephson current $\sin (\varphi_l)$ (third row) without the in-plane dissipation $\beta_{ab}=0$ (a-d) and with the in-plane dissipation $\beta_{ab}=0.2$ (e-f). When $\beta_{ab}=0$, various cavity modes $(m, n)=(7,5)$ in (a), $(m, n)=(9,3)$ in (b), $(m, n)=(7,2)$ in (c) and $(m, n)=(4,1)$ in (d)  are excited when the bias current is swept. For $\beta_{ab}=0.2$, only the modes with $n=0$, [$(m, n)=(1,0)$ in (e), other cavity modes with $m>1$ are not shown here] are excited for $N=10^3$. For $N=10^4$, irregular patterns of the EM fields are excited even with the in-plane dissipation $\beta_{ab}=0.2$. The supercurrent forms blocks with alternating sign between neighbouring blocks indicating $\pm\pi$ phase jumps at the interface of each block. We use $L_x=0.4\lambda_c$, $\zeta=7.1\times 10^4$ in simulations.}
\end{figure*}

For the solution that is homogeneous along the $x$-axis but non-uniform along the $c$-axis, the radiation contribution can be neglected and the phase oscillates according to $\varphi_l=\phi_l+\omega t +i/(-\omega^2+i\beta\omega)$, where $\phi_l$ accounts for the phase shifts in different junctions and is randomly distributed. The stability analysis is the same as the case of uniform solution, but now there is no radiation contribution $R_r=R_i=0$. The stability diagram is the same as that in Fig. \ref{f1} because the radiation contribution is small for $L_x\approx 80 \rm{\ \mu m}$ used in experiments.

For a small crystal proposed in Ref. \onlinecite{Bulaevskii07}, $L_x\approx 4\rm{\ \mu m}$, $\omega_c\approx 150$. The cavity modes cannot be excited for such a high frequency according to Eq. (\ref{eq19}). The homogeneous solution along the lateral directions is thus stable. In the single junction limit $N=1$, the parametric instability leads to the excitation of solitons, which manifests as the zero-field current step in \emph{IV} characteristics\cite{Pagano86}.

For non-uniform perturbations along the $c$ axis, the in-plane current $J_{x,l}$ is induced according to the Ampere's law $4\pi J_{x,l}/c=-\left(B_{y,l+1}-B_{y,l}\right)/s$ where we have neglected the displacement current \cite{Bulaevskii94}. $J_{x,l}$ has contribution from the normal current and
supercurrent $J_{x,l}=\sigma _{\text{ab}}\frac{\Phi _0}{2\pi  c}\partial _t\mathcal{P}_l+\frac{c \Phi _0}{8\pi ^2\lambda _{\text{ab}}^2}\mathcal{P}_l$ where $\mathcal{P}_l=\partial _x\Theta _l-2\pi  A_x/\Phi _0$ is the in-plane superconducting momentum with superconducting phase $\Theta_l$ and vector potential $A_x$. The in-plane dissipation in units of $\frac{\omega _J}{4\pi } \left(\frac{\Phi _0}{2\pi  \lambda _c s}\right)^2$ is
\begin{equation}\label{eq20}
P_{ab}(k_m, q,\omega )=\frac{2\sin ^2(q/2)\beta _{\text{ab}}\omega ^2}{\left(\beta _{\text{ab}} \omega \right)^2+1}\left|c_q^4\right| k_m^2\left|a_p\left(k_m,q\right)\right|^2,
\end{equation}
which is much larger than the dissipation along the $c$-axis $P_{\text{c}}(k_m, q,\omega )=\frac{1}{2}\omega ^2\beta _c\left|a_p\left(k_m,q\right)\right|^2$ for $q$'s when ${\sin ^2(q/2)>[{\left(\beta _{\text{ab}} \omega \right)^2+1}]/[4\beta _{\text{ab}}}\left|c_q^4\right| k_m^2]$. Thus the non-uniform perturbations with a large $q$ quickly die out due to the strong in-plane dissipation $P_{ab}$. While for perturbations with $q\ll 1$, the in-plane dissipation is weak or absent, and the perturbations lead to the excitation of the cavity mode with a small $q$.

 \begin{figure}[b]
\psfig{figure=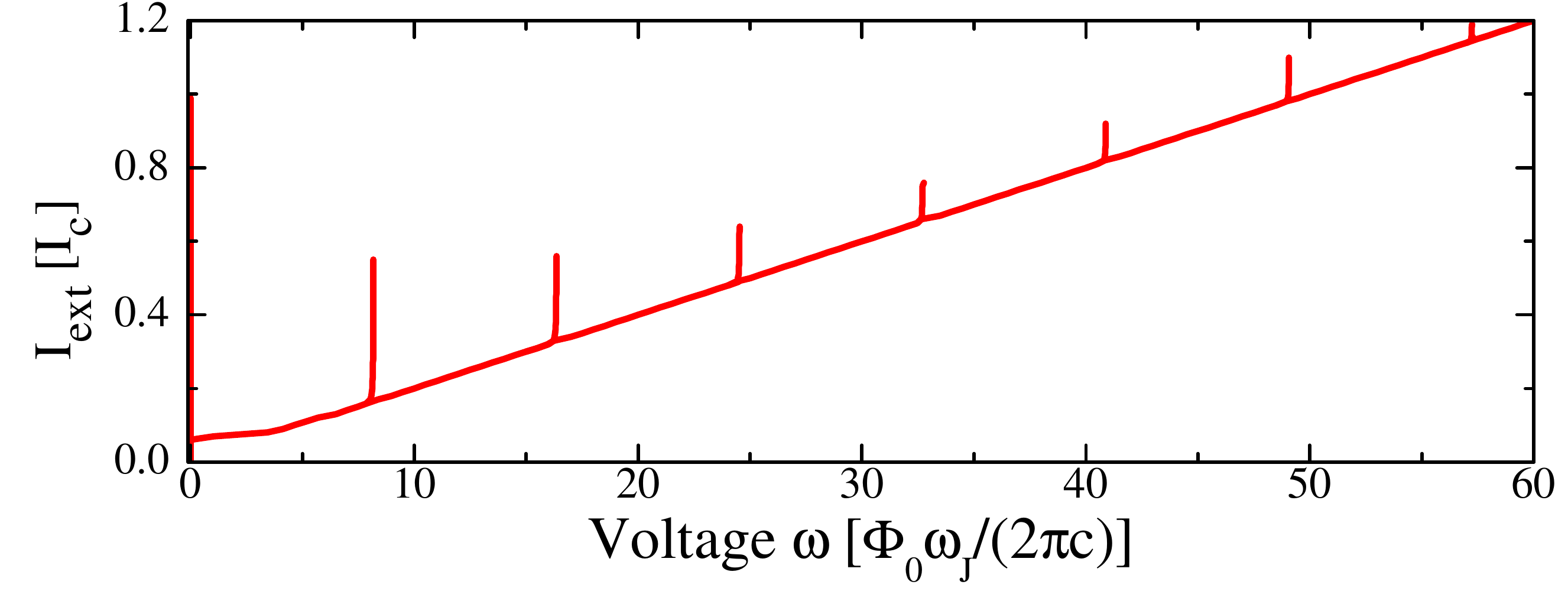,width=\columnwidth}
\caption{\label{f3}(color online) \emph{IV} curve obtained with $\beta_{ab}=0.2$ and $N=10^3$. Other parameters are the same as those in Fig. \ref{f2}. The radiation power is estimated using $S_r=|E_{ac}^2|/(2|Z|)$ and $\epsilon_d=1$ where $E_{ac}$ is the averaged ac electric field at edges\cite{szlin09a}. The maximal power at the first cavity mode $m=1$ is about $2800\rm{W/cm^2}$. The experimental measured \emph{IV} deviates significantly from the theoretical one in high bias region due to the strong self-heating effect.}
\end{figure}

\section { Numerical simulation } 
We also perform numerical simulations by solving Eqs. (\ref{eq1}) and (\ref{eq2}) with the non-radiation boundary condition $B_{y,l}=\pm I_{\rm{ext}} L_x/2$ to check the above analytical results. For numerical details, see Appendix A. Let us first present the results without in-plane dissipation $\beta_{ab}=0$ as shown in Fig. \ref{f2}(a-d). Upon sweeping the current, various cavity modes both along the $x$-axis and $c$-axis are excited for $N=10^3$. It is difficult to excite the uniform mode $q=0$ due to the existence of other competing cavity modes in the crystal with this large $N$ at a given voltage. Interestingly, when we turn on the in-plane dissipation by putting $\beta_{ab}=0.2$, only the cavity modes uniform along the $c$-axis can be excited when the current is swept as shown in Fig. \ref{f2}(e), consistent with the above analytical results. The corresponding \emph{IV} characteristics is shown in Fig. \ref{f3}. At the cavity resonances, a large amount of energy is pumped into the plasma oscillation and current steps are induced, and the radiation power is enhanced. 

For $N\lesssim 4000$ in simulations, only the uniform mode can be excited due to the in-plane dissipation, consistent with the results in Fig. \ref{f1}. For even larger number of IJJs, such as $N=10^4$, irregular patterns of EM fields are developed in simulations for the adopted parameters, see Fig. \ref{f2}(f). It is consistent with the analytic result summarized in Fig. \ref{f1} where cavity modes with $n=0, 1, 2$ can be excited. Other mechanisms to synchronize all junctions are needed in order to achieve strong radiation for thick mesas. Here, we note that for small values of $N<100$, the voltage corresponding to various cavity modes is discrete because $c_q$ is well separated for different values of $n$, thus one can select the cavity modes simply by tuning the voltage.

For $\beta_{ab}=1.0$, we found numerically that the uniform plasma oscillation with $q=0$ becomes stable again for $N=10^4$. This suggests that stronger in-plane dissipation is important to achieve uniform plasma oscillation. $\beta_{ab}$ increases with temperature, while on the other hand thermal fluctuations tend to destroy the uniform oscillation. Therefore the synchronization is optimal at some intermediate temperatures.

One peculiar feature in Fig. \ref{f2} is that supercurrent forms blocks in space, where the current change sign between neighboring blocks, or equivalently at nodes of the oscillating electric field. This means that there is $\pm \pi$ phase jump or $\pm \pi$ phase kink at the interface of blocks. The phase kinks stack along the $c$-axis with alternating signs, such as $(\cdots,1, -1, 1, -1, \cdots)\pi$. The state with phase kink with $q=0$ was first suggested in Refs. \onlinecite{szlin08b,Koshelev08b} as a possible mechanism for strong THz radiation observed in experiments. A characterization of the kink sate with a general $q$ is presented in Appendix B.

 \section{ Discussions} One should also check the stability of the excited cavity modes. This has already been done in Refs.\onlinecite{szlin10pc,Koshelev10,szlin09a} and found that the kink states associated with cavity modes with $q=0$ are stable for a small $N$, which is also confirmed by the results in Fig. \ref{f2}. For a large $N$, it was shown in Ref. \onlinecite{Koshelev10} that long wavelength instability develops and the kink states with uniform plasma oscillation $q=0$ becomes unstable. This is consistent with the results in Fig. \ref{f2}(f).

When a strong magnetic field is applied parallel to the $ab$-plane, the Josephson vortices (JVs) are induced. The JVs favor the triangular lattice in low velocity region due to the strong inter-vortex repulsion in a long IJJs stack \cite{Krasnov1999}. As shown in Appendix C, our simulations show that the in-plane dissipation mechanism becomes insufficient to achieve the rectangular JVs lattice and in-phase plasma oscillation, consistent with the analytical results in Ref. \onlinecite{Koshelev01}. A new mechanism for synchronization is needed in this case.

 \section { Acknowledgements} 
 The authors are grateful to L. N. Bulaevskii for critical reading of the manuscript and helpful discussions. SZL gratefully acknowledges funding support from the Office of Naval Research via the Applied Electrodynamics collaboration. XH is supported by WPI Initiative on Materials Nanoarchitectonics, MEXT of Japan, and by CREST, JST.

\appendix

\section{Numerical method}
\begin{figure}[b]
\psfig{figure=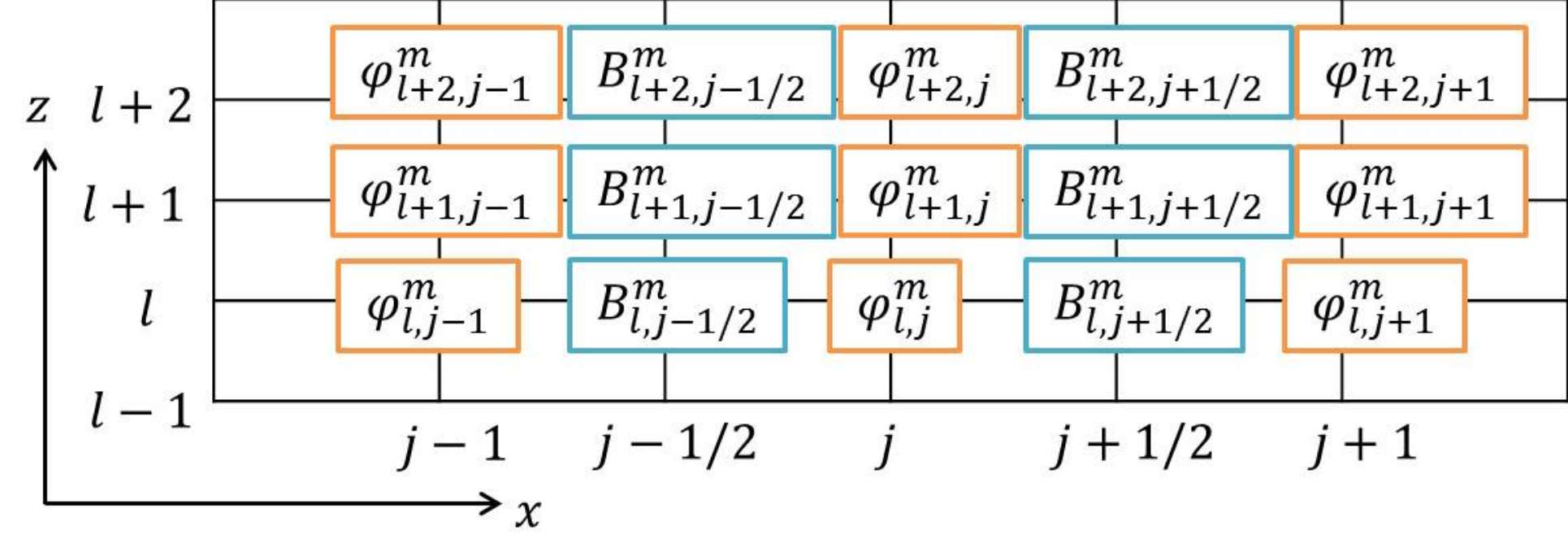,width=\columnwidth}
\caption{\label{fs1}(color online) Schematic view of the numerical grids.}
\end{figure}

Here we present the numerical method to solve Eqs. (\ref{eq1}) and (\ref{eq2}). The model is discrete inherently along the $c$-axis, and we only need to discretize along the $x$-axis. The phase $\varphi$ is defined at nodes $(j,l)$ and magnetic field $B_y$ is defined at nodes $(j+1/2, l)$ with integer $j$ and $l$, see Fig. \ref{fs1}. The grid size along $x$ is $dx$ and time step is $dt$. Equation (\ref{eq1}) then becomes
\begin{align}\label{eqan1}
\nonumber \frac{\varphi _{l,j}^{m+1}+\varphi _{l,j}^{m-1}-2\varphi _{l,j}^m}{dt^2}+\beta _c\frac{\varphi _{l,j}^{m+1}-\varphi _{l,j}^{m-1}}{2dt}+\sin\varphi _{l,j}^m\\
=\frac{B_{l,j+1/2}^m-B_{l,j-1/2}^m}{dx}.
\end{align}
We know $\varphi$ and $B_y$ at the $m$-th time step and we can obtain $\varphi$ at the $(m+1)$-th step directly from Eq. (\ref{eqan1}). We use an implicit method to discretize Eq. (\ref{eq2}). After some simple manipulations, we have
\begin{align}\label{eqan2}
\nonumber \frac{{\varphi _{l,j + 1}^{m + 1} - \varphi _{l,j - 1}^{m + 1} + \varphi _{l,j + 1}^m - \varphi _{l,j - 1}^m}}{{2dx}} \\
\nonumber + {\beta _{{\rm{ab}}}}\frac{{\left( {\varphi _{l,j + 1}^{m + 1} - \varphi _{l,j - 1}^{m + 1}} \right) - \left( {\varphi _{l,j + 1}^m - \varphi _{l,j - 1}^m} \right)}}{{dtdx}} \\
+ {\beta _{{\rm{ab}}}}\frac{{B_{l,j + 1/2}^m}}{{dt}} - \left( {1 - \zeta {\Delta ^{(2)}}} \nonumber \right)\frac{{B_{l,j + 1/2}^m}}{2}\\
= \left( {1 - \zeta {\Delta ^{(2)}}} \right)\frac{{B_{l,j + 1/2}^{m + 1}}}{2} + {\beta _{{\rm{ab}}}}\frac{{B_{l,j + 1/2}^{m + 1}}}{{dt}}.
\end{align}
Equation (\ref{eqan2}) can be written as a matrix equation. Inverting the matrix at the right-hand side of Eq. (\ref{eqan2}) using $\varphi_{l,j}^{m+1}$ obtained from Eq. (\ref{eqan1}), we then obtain $B_{l,j+1/2}^{m+1}$. The electric field is given by $E_{l, j}^m=({\varphi _{l,j}^{m+1}-\varphi _{l,j}^{m-1}})/({2dt})$.

\begin{figure*}[t]
\psfig{figure=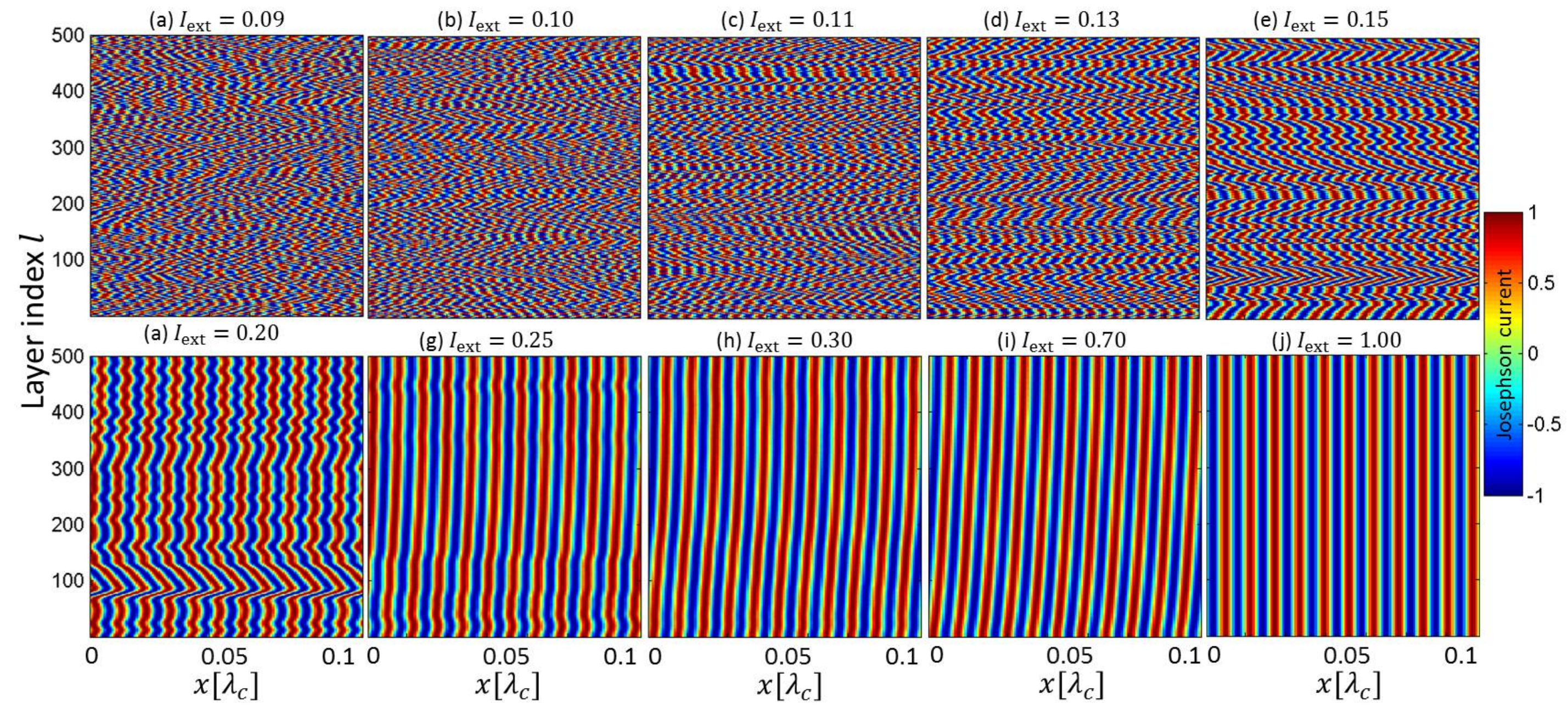,width=18cm}
\caption{\label{fs2}(color online) Snapshots of the Josephson current $\sin (\varphi_l)$ in the flux-flow region with the in-plane dissipation $\beta_{ab}=0.2$. Here the applied magnetic field is $B_a=1\ \rm{T}$, $L_x=0.1\lambda_c$, $N=500$, $\zeta=7.1\times 10^4$. The core of Josephson vortex is located at $\varphi=(2m+1)\pi$.}
\end{figure*}

\section{Characterization of the kink state with a general $q$}

In the kink state, $\varphi_l$ can be written as
\begin{equation}\label{eq21}
\varphi _l=\omega  t+\varphi _{s,l}(x)-i \sum_{m,q}A_{m,q}\cos(k_m x)\cos( q l) \exp (i \omega  t),
\end{equation}
where the rotating phase at the right-hand side (rhs) of Eq. (\ref{eq21}) is due to the voltage, the second term at rhs is the static phase kink and the last term at rhs is the cavity mode both along the $c$-axis and $x$-axis. For stacks with a large $N$, there are many cavity modes $(m, q)$ corresponding to a same frequency $\omega$, thus a summation over all modes are needed \cite{szlin09a}. The magnetic field is
\begin{equation}\label{eq22}
B_l=B_{s,l}(x)-i \sum_{m,q}C_{m, q} \sin(k_m x)\cos( q l) \exp (i \omega  t),
\end{equation}
where $B_{s,l}$ is the static magnetic field due to the phase kink. We have neglected the frequency harmonics in Eqs. (\ref{eq21}) and (\ref{eq22}), which is valid when $A_{m,q}<1$. From Eq. (\ref{eq2}) we obtain $C_{m,q}=-A_{m,q} k_m c_q^2$ and $[1-\zeta \Delta^{(2)}]B_{s,l}(x)=\partial_x \varphi _{s,l}(x)$.
Substituting Eqs. (\ref{eq21}) and (\ref{eq22}) into Eq. (\ref{eq1}), we obtain a closed equation for $\varphi_l$. For the frequency component $\omega$, we have
\begin{equation}\label{eq25}
\left[i k_m^2c_q^2-\left(\beta_c  \omega +i \omega ^2\right)\right]A_{m,q} \cos \left(k_mx\right)\cos \left( q l\right)=-i e^{ i \varphi _{s,l}}.
\end{equation}
Projecting $\exp (i \varphi _s)$ into the cavity mode $(m, q)$, we obtain the amplitude of the plasma oscillation
\begin{equation}\label{eq26}
A_{m,q}=\frac{-i F_{m,q} }{i k_m^2c_q^2-\left(\beta_c  \omega +i \omega ^2\right)},
\end{equation}
with the coupling between the cavity mode and phase kink
\begin{equation}\label{eq27}
F_{m,q} =\frac{\alpha}{N L_x}\sum_{l=0}^N\int \exp (i \varphi _{s,l}) \cos \left(k_m x\right)\cos \left(q l\right)d x,
\end{equation}
where $\alpha=2$ for $q=0$ and $\alpha=4$ for $q>0$. When the voltage is tuned close to the cavity resonance, the amplitude of plasma oscillation is enhanced. The linewidth of the resonance is determined by $\beta_{ab}$ and $\beta_c$. The \emph{IV} characteristics is given by $I_{\rm{ext}}=\beta_c\omega+\left\langle\sin(\varphi_l)\right\rangle_{x,l,t}$ where $\left\langle\cdots\right\rangle_{x,l,t}$ denotes average over space and time. We then obtain
\begin{equation}\label{eq27a}
I_{\rm{ext}}=\beta_c\omega+ {\rm{Re}}\left[\sum_{m,q}\frac{|F_{m,q}|^2/(2\alpha)}{i k_m^2c_q^2-\left(\beta_c  \omega +i \omega ^2\right)}\right].
\end{equation}

Now let us consider the static component
\begin{equation}\label{eq28}
\partial _x^2\varphi _s(x, z)=\frac{i \zeta}{2}  \Delta ^{(2)}\sum_{m,q} \left[A_{m,q}{\cos(k_m x)\cos( q l)}e^{-i \varphi _{s,l}} \right].
\end{equation}
The solution of $\varphi_{s,l}$ depends on the spatial profile of the plasma oscillation. To present the analytical results, we consider a case that the plasma profile has well-defined nodes as shown in Fig. \ref{f2} (a-d), where we may approximate the spatial profile by a dominant mode $A\cos(k_1 x)\cos( q l)$. Without loss of generality, we have taken the $m=1$ mode.  The variation of $\varphi_{s,l}$ along the $c$-axis is $\varphi_{s,l}=(-1)^l\varphi_{s0}$ except for the node region of the plasma oscillation, which is much faster than the plasma mode $q$. We may approximate Eq. (\ref{eq28}) as
\begin{equation}\label{eq29}
\partial _x^2\varphi _{s0}=2{ \zeta {\rm{Re}}[A]} {\cos(k_1 x)\cos( q l)}\sin (\varphi _{s0}).
\end{equation}
Equation (\ref{eq29}) is invariant under the transformation $x\leftarrow L_x-x$ and $\varphi_{s0}\leftarrow\pi-\varphi_{s0}$, which gives the static $\pi$ phase kink along the $x$-axis at the nodes of oscillating electric field when $\cos(k_1 x)=0$. The width of the kink is $\lambda_k=1/\sqrt{2\zeta {\rm{Re}}[A]|\cos(ql)| }$, which is small $\lambda_k\ll 1$ except for the node region of $\cos(q l)\approx 0$. The kink in the $l$-th junction can be excited only when $\lambda_k\ll L_x$. Near the nodes of the oscillating $E_z$ or $B_y$ along the $c$-axis where $\cos(q l)\approx 0$, $\lambda_k$ may be comparable to $L_x$ thus no kink exists in the node region, consistent with results in Fig. \ref{f2}. When $\cos(q l)$ changes sign, $\varphi_{s0}$ acquires a $\pi$ shift, because Eq. (\ref{eq29}) is invariant when $\cos(ql)\leftarrow -\cos(ql)$ and $\varphi_{s0}\leftarrow\pi+\varphi_{s0}$. Thus there are $\pm\pi$ phase jumps at the nodes of oscillating electric fields both along the $c$ and $x$ axis.

\section{Effect of the in-plane dissipation on the dynamics of Josephson vortices}
When a strong magnetic field is applied perpendicularly to the $c$-axis of BSCCO single crystal, the Josephson vortices are induced and form the triangular lattice. Driving by the Lorentz force induced by a transport current, the Josephson vortices move and excite Josephson plasma. The motion of Josephson vortex lattice provides an alternative routine to achieve a strong THz radiation. \cite{Bae07} Due to the strong inter-vortex repulsion, the Josephson vortices favor the triangular lattice\cite{Kim05,Krasnov99b} and the radiation is weak. The rectangular lattice is observed in a small mesa\cite{Katterwe09} where the surface potential favors the rectangular lattice. Here we investigate the possible synchronization of the Josephson vortices by the in-plane dissipation.

The simulation results are presented in Fig. \ref{fs2}. When the bias current increases, the Josephson vortices evolve toward a rectangular lattice. The rectangular lattice is achieved only by a high bias current, which is difficult to realize experimentally due to the strong self-heating effect. Our simulations are consistent with the analytical results obtained by Koshelev and Aranson\cite{Koshelev01}, who found that the rectangular lattice can only be stability in a high velocity region. The in-plane dissipation does not stabilize the in-phase plasma oscillation or rectangular lattice of Josephson vortices due to the strong inter-vortex repulsion. The realization of the in-phase oscillation in the case of Josephson vortices is still an open problem and requires a new mechanism.


\begin{thebibliography}{35}
\bibitem [{uni()}]{units}%
  \BibitemOpen
  \href@noop {} {}\bibinfo {note} {The inductive coupling is denfined as
  $\zeta=(\lambda_{ab}/s)^2$ and the renormalized conductivities along the $c$
  axis and $ab$ plane are defined as $\beta_c=4\pi\sigma_c/(\epsilon_c
  \omega_J)$ and $\beta_{ab}=4\pi
  \sigma_{ab}\lambda_{ab}^2/(\lambda_c^2\epsilon_c\omega_J)$ with the Josephson
  plasma frequency $\omega_J=c/(\lambda_c\sqrt{\epsilon_c})$. Here
  $\lambda_{ab}$ and $\lambda_c$ are the London penetration depths and $s$ is
  the period of the stack of IJJs. In Eqs. (\ref{eq1}) and (\ref{eq2}),
  frequency is in units of $\omega_J$, length is in units of $\lambda_c$ and
  magnetic field is in units of $\Phi_0/(2\pi\lambda_c s)$ with
  $\Phi_0=hc/(2e)$ the flux quantum. For BSCCO, $\omega_J/(2\pi)\approx 0.1
  \rm{\ THz}$, $\lambda_{ab}\approx 0.4\rm{\ \mu m}$ and $\lambda_c\approx 200
  \rm {\ \mu m}$. The dimensionless electric field is given by
  $E_{z,l}=\partial_t\varphi_l$, with $E$ in units of $\Phi_0\omega_J/(2\pi c
  s)$.}\BibitemShut {Stop}%
\end{thebibliography}
%

\end{document}